\documentclass[12pt]{iopart}

\usepackage{iopams}
\usepackage{graphics}
\usepackage{graphicx}
\usepackage{epsfig}
\begin{document}
\title{Bose-Einstein condensation and entanglement in magnetic systems}
\author{M. A. Continentino}

\address{Instituto de F\'{\i}sica,
Universidade Federal Fluminense, Campus da Praia Vermelha,
Niter\'{o}i, 24210-340, RJ, Brazil}
\ead{mucio@if.uff.br}
%\date{\today}

\begin{abstract}
We present a study of magnetic field induced quantum phase
transitions in insulating systems. A generalized scaling theory is
used to obtain the temperature dependence of several physical
quantities along the quantum critical trajectory ($H=H_{C}$,
$T\rightarrow0$) where $H$ is a longitudinal external magnetic
field and $H_{C}$ the critical value at which the transition
occurs. We consider transitions from a spin liquid at a critical
field $H_{C1}$ and from a fully polarized paramagnet, at $H_{C2}$,
into phases with long range order in the transverse components.
The transitions at $H_{C1}$ and $H_{C2}$ can be viewed as
Bose-Einstein condensations of magnons which   however belong to
different universality classes since they have different values of
the dynamic critical exponent. Finally, we use that the magnetic
susceptibility is an entanglement witness to discuss how this type
of correlation sets in as the system approaches the quantum
critical point along the critical trajectory, $H=H_{C2}$,
$T\rightarrow0$.

\end{abstract}
\maketitle

\section{Introduction}

Recently there has been an intense study of field induced quantum
phase transitions in metallic \cite{fi1,fi2,fi3,fi4,fi5,fi6} and
insulating materials \cite{i1,i2,i3,i4,i5,i6}. These field induced
transitions are generally associated with a Bose-Einstein
condensation of magnons. In the insulating case, most of the
experimental studies, have concentrated in obtaining the
\emph{shift exponent} that characterizes the shape of the critical
line in the neighborhood of the quantum critical point (QCP).
However, it is important to have additional experimental
information that can be used to fully characterize the
universality classes of the different zero temperature field
induced transitions. For this purpose it is convenient to obtain
the thermodynamic behavior of the system along a special
trajectory in the phase diagram. This \textit{quantum critical
trajectory} consists essentially in \textit{sitting} at the
quantum critical point, by fixing the magnetic field at its
critical value and vary the temperature. Field induced transitions
are generally associated with soft modes. Then, at the QCP where
the gap for excitation vanishes, physical quantities as specific
heat, magnetization and susceptibility have power law temperature
dependencies determined by the quantum critical exponents. Our aim
here is to obtain this behavior. This strategy has been
intensively explored in the study of heavy fermion materials, in
this case, fixing the pressure at its critical value for the
disappearance of magnetic order \cite{stewart}. This approach is
particularly useful for insulating materials where the magnons
that become soft at the QCP account for most of the low
temperature thermodynamic behavior. In metals, even in field
induced transitions, one has to consider also the contribution of
charge carrying excitations \cite{fi1,fi2,fi3,fi4,fi5,fi6}.  The
soft magnon modes couple to these excitations and both may be
strongly renormalized.

We will consider two types of field induced transitions. Firstly,
we study the zero temperature transition in an antiferromagnet
system, from the fully polarized paramagnetic state into a phase
with transverse components of the magnetization as the
longitudinal magnetic field is \emph{reduced} to a critical value
$H_{C2}$  \cite{mucioRC}. This is a second order transition which
is characterized by a dynamic critical exponent $z=2$
\cite{mucioRC}. It can be identified with a Bose-Einstein
condensation of magnons with the magnetic field playing the role
of the chemical potential. We also consider the transition from a
disordered spin-liquid phase to an antiferromagnetic phase
\cite{i5}, at a critical magnetic field, $H_{C1}$. In this case
the field is increased to the critical value $H_{C1}$ at which the
the singlet gap vanishes and magnetic long range order sets in.
This transition is characterized by a dynamic critical exponent
$z=1$ \cite{crisan}. In both cases the transition is approached
from the \emph{disordered} side, i.e., with $H>H_{C2}$ and
$H<H_{C1}$. Although these transitions have been intensively
studied \cite{i3,i4,i5,16,sachdev} the physical behavior along the
quantum critical trajectory, $H=H_{C},T\rightarrow0$ has not been
sufficiently characterized. In this paper we fully determine the
thermodynamic properties and the appearance of entanglement along
this line. We hope in this way to motivate further experiments
that go beyond obtaining the shift exponent of the critical line.
Notice that away from the QCP, in the disordered phase, the
excitations are gapped and the physical behavior is thermally
activated.

Our results are obtained using a generalized quantum scaling
theory which was previously applied to heavy fermion materials
\cite{mucio,livro}. This theory describes completely the
thermodynamic behavior along the quantum critical trajectory. The
reason is that we consider three dimensional materials and the
effective dimensions \cite{mucio,livro} at the field induced
quantum phase transitions $d_{eff}=d+z$ are at, or above, the
upper critical dimension $d_{c}=4$ for these transitions. In this
case, the quantum critical exponents assume Gaussian or mean-field
values, being fully determined. Logarithmic corrections which
arise in the case $d_{eff}=d_{c}$ are not obtained in the present
approach. The microscopic justification of the scaling theory is
provided by renormalization group calculations \cite{millis,
sachdev}. In the last section we discuss how entanglement sets in
among the spins along the quantum critical trajectory.

\section{Scaling analysis of the quantum phase transition}

We consider first an antiferromagnetic system with longitudinal
anisotropy and a strong magnetic field applied along the easy axis
direction. We study the $T=0$ transition, with decreasing magnetic
field, from the fully polarized paramagnet to a phase with long
range order in the transverse components of the magnetization at a
critical field $H_{C2}(T=0)$ \cite{mucioRC}. This problem can be
treated as a dilute gas of bosons with the effective action given
by \cite{sachdev},
\begin{equation}
\label{uzunov}
S=\frac{1}{2}\int_{\mathbf{k},\omega}[i\omega+Dk^{2}+\delta]|\psi
(\mathbf{k},\omega)|^{2}+v_{0}\int_{\mathbf{x},\tau}|\psi|^{4}%
\end{equation}
where $\delta=H-H_{C2}$. The field $\psi(\mathbf{k},\omega)$ is a
two-component field representing the components of the spins
transverse to the direction of the magnetic field and $v_{0}$
takes into account the spin-wave interactions. The transition at
$\delta=0$ has a dynamic exponent $z=2$ due to the
ferromagnetic-like dispersion of the magnons \cite{mucioRC}. This
action has been extensively studied in the context of the
non-ideal Bose gas \cite{uzunov}, the dilute Bose gas
\cite{fisherd} and superfluid-insulator transitions
\cite{fisherm}. In our case it is useful  due to the small number
of magnons excited at the field induced quantum phase transition
\cite{kawa}. It takes into account the ferromagnetic-like
dispersion of these modes \cite{mucioRC} and incorporates a
constraint in their total number \cite{kawa}.

The thermodynamic properties of systems, close to the QCP
described by the action in \Eref{uzunov}, can be obtained from the
free energy density. This has the scaling form \cite{livrop},
\begin{equation}
f\propto|\delta(T)|^{2-\alpha}F\left(  \frac{T}{|\delta(T)|^{\nu z}}\right)
\label{gaussian}%
\end{equation}
near the QCP. The zero temperature critical exponents $\alpha$,
$\nu$ and the dynamic exponent $z$ are related to the
dimensionality of the system $d$ by the quantum hyperscaling
relation, $2-\alpha=\nu(d+z)$ \cite{livro}. In general for
$d_{eff}=d+z>4$, i.e., above the upper critical dimension
$d_{c}=4$, the exponents associated with the QCP at $\delta=0$
take Gaussian values and in particular the correlation length
exponent, $\nu=1/2$. For the action in \Eref{uzunov} these
exponents remains Gaussian even below $d_c=2$ \cite{uzunov}.
Although within the renormalization group approach for $d=3$, the
transition at $\delta=0$ is controlled by the Gaussian fixed
point, the spin-wave coupling $v_{0}$ acts as a \emph{dangerously}
irrelevant interaction for $d_{eff}>4$ and must be dealt with
carefully \cite{millis}. Perturbation theory in powers of $v_{0}$
leads to a temperature dependent critical line given by,
$\delta(T)=H_{C2}(T)-H_{C2}+v_{0}T^{1/\psi}=0$ with the shift
exponent $\psi=z/(d+z-2)=2/3$ in three dimensions \cite{millis}.

In order to obtain the correct scaling behavior near the QCP, the
scaling function $F(x)$ in Eq. \ref{gaussian} must have the
asymptotic behaviors,
\begin{equation}
\label{cases}
F(x)=\cases{constant&for $x \rightarrow 0$\\
x^p&for $x \rightarrow \infty$\\}
\end{equation}
The first guarantees that we recover the correct behavior at
$T=0$, with
$f\propto|\delta|^{2-\alpha}$. The second with $p=(\widetilde{\alpha}%
-\alpha)/\nu z$, yields
\begin{equation}
\label{tilde}
f(T)\propto A(T)|\delta(T)|^{2-\tilde{\alpha}}%
\end{equation}
where $A(T)=T^{p}$ \cite{livrop}. \textit{Tilde} exponents refer
to finite temperature transitions and $\tilde{\alpha}$ is the
Gaussian thermal specific heat exponent of the $3d-XY$ model. This
is related to the thermal Gaussian correlation length exponent
$\tilde{\nu}$ through the Josephson relation,
$2-\tilde{\alpha}=\tilde {\nu}d$. Since $\nu=\tilde{\nu}=1/2$ for
$d+z>4$, the free energy in the neighborhood of the QCP can be
written as, $f\propto T|\delta(T)|^{\tilde{\nu }d}$ ($p=1$). From
this expression we obtain the temperature dependence of the
magnetization, susceptibility and specific heat at $H=H_{C2}$,
\numparts
\begin{eqnarray}
m&=a_{1}T^{3/2}+b_{1}v_{0}^{1/2}T^{7/4}\\
\chi&=a_{2}T^{1/2}+b_{2}(1/v_{0})^{1/2}T^{1/4}\\
C_{V}&=a_{3}T^{3/2}+b_{3}v_{0}^{3/2}T^{9/4}
\end{eqnarray}
\endnumparts
respectively, where the $(a_{i},b_{i})$ are constants. For each
quantity, the first term is the Gaussian contribution and the
second arises from corrections due to the dangerous irrelevant
spin-wave interaction $v_0$. Notice that, except for the
susceptibility and correlation length (see Table \ref{table}), for
which $v_{0}$ behaves as a truly dangerously irrelevant
interaction as it appears in a denominator, the purely Gaussian
term is dominant at low temperatures. A similar analysis can be
carried out for the transition from the spin liquid to the
antiferromagnet at $H_{C1}$ with a dynamic exponent $z=1$. The
results are given in Table \ref{table}. This case is {\em marginal}, since $d_{eff}%
=d_{c}=4$ is the upper critical dimension and there are logarithmic
corrections to the quantities in Table \ref{table}. These however are not
obtained in the scaling approach (see Refs. \cite{crisan, MCQ}).

\begin{table}[ptbh]
\begin{center}%
\begin{tabular}
[c]{|c|c|c|c|}\hline
Physical Quantity &  & $H_{C1}(z=1)$ & $H_{C2}(z=2)$\\\hline
Shift exponent & $\psi=\frac{z}{d+z-2}$ & $1/2$ & $2/3$\\\hline
Magnetization & $-\partial f/\partial H$ & $T^{2}$ & $T^{3/2}$\\\hline
Susceptibility & $-\partial^{2}f/\partial H^{2}$ & $\frac{1}{\sqrt{v_{0}}}$ &
$\frac{1}{\sqrt{v_{0}}}T^{1/4}$\\\hline
Specific heat & $-T\partial^{2}f/\partial T^{2}$ & $T^{3}$ & $T^{3/2}$\\\hline
Correlation length & $\frac{1}{\sqrt{v_{0}}}T^{-\nu/\psi}$ & $\frac{1}%
{\sqrt{v_{0}}}T^{-1}$ & $\frac{1}{\sqrt{v_{0}}}T^{-3/4}$\\\hline
\end{tabular}
\end{center}
\caption{Power law temperature dependence of physical quantities along the
critical trajectory ($H=H_{C}$, $T\rightarrow0$). Logarithmic corrections are
not included. Only the dominant contribution in the low T limit is given
\cite{roschnote}.}%
\label{table}%
\end{table}

Away from the QCP, for $H>H_{C2}$ and $H<H_{C1}$, there are gaps for
excitation of spin-waves and the crossover temperature $T_{\times}%
\propto|\delta|^{\nu z}$ gives the energy scale for the thermally
activated behavior of the thermodynamic functions in these regions
of the phase diagram. Since these gaps vanish as $|\delta|^{\nu
z}$ their measurement allows for a determination of the gap
exponent $\nu z$. Notice that the difference in universality
classes of the two transitions studied above arises essentially
from the different dispersion relations of the soft magnons at the
QCP.

Finally at $T=0$ in the ordered phase, it is also necessary to take
into account the dangerous irrelevant spin-wave interaction $v_0$.
The scaling form of the free energy is \cite{livro}, $f \propto
|\delta|^{\nu(d+z)}F_v\left[ v_0|\delta|^{(d+z-4)/2} \right]$. The
dangerous irrelevant nature of $v_0$ is manifested in the fact that
the scaling function $F_v[x \rightarrow 0]\propto 1/x$. This yields
\[
f \propto \frac{|\delta|^{\nu(d+z)}}{v_0|\delta|^{(d+z-4)/2}} =
\frac{|\delta|^2}{v_0}
\]
for $d+z>4$. This mean-field behavior gives rise to a {\em
longitudinal} magnetization varying linearly with the distance to
the QCP, i.e., $M \propto |H-H_C|$.

\section{Entanglement and the quantum phase transition}

Recently, there has been a large interest in characterizing
entanglement in systems near quantum critical points and in
macroscopic magnetic systems \cite{claim}. For a system with N
spins, Wiesniak et al. \cite{vedral} have shown that the magnetic
susceptibility acts as an entanglement witness and that whenever,
\begin{equation}
\tilde{\chi}=\chi_{x}+\chi_{y}+\chi_{z}\leq\frac{N\mathit{l}}{kT}%
\label{entanglement}%
\end{equation}
there is entanglement between the individual spins of magnitude $\mathit{l}$.
The $\chi_{i}$ are the susceptibilities along three orthogonal axis measured
in the same quantum state. A quantum complementarity relation involving the
susceptibility and magnetization can also be obtained \cite{vedral} ($T\neq
0$),
\begin{equation}
1-\frac{kT\tilde{\chi}}{N\mathit{l}}+\frac{M^{2}}{N^{2}{\mathit{l}}^{2}}%
\leq1\label{complementarity}%
\end{equation}
This is particularly useful when applied to low dimensional
materials, or systems at quantum criticality, as temperature can
be reduced without any phase transition. In the equation above, it
is useful to define the quantity
$E(T,H)\equiv1-(kT\tilde{\chi})/N\mathit{l}$ which provides a
measurement of entanglement, while the last term $S\equiv
M^{2}/N^{2}{\mathit{l}}^{2}$ represents local properties. Equation
\ref{complementarity} shows the interplay between these quantities
since, $0\leq E+S\leq1$.

It is interesting to apply the relations above to the previous
problems. Let us consider the case $z=2$, with the applied
magnetic field fixed at the critical value $H_{C2}$. For $T=0$ and
$H\geq H_{C2}$ the system is in a fully polarized state,
$M=N\mathit{l}$, and we must have, $E(T\rightarrow
0,H=H_{C2})\rightarrow0$. This implies that, as $T\rightarrow0$,
$\tilde{\chi }=\chi_{x}+\chi_{y}+\chi_{z}=N\mathit{l}/kT$. Since
the system is already fully polarized at $T=0$, we must have
$\chi_{z}(T\rightarrow0,$ $H=H_{C2})\rightarrow0$ and
consequently, $\chi_{x}(T\rightarrow
0,H=H_{C2})=\chi_{y}(T\rightarrow0,H=H_{C2})=N\mathit{l}/2kT$.
Assuming that the transverse uniform susceptibilities have already
taken their low temperature asymptotic behavior as entanglement
sets in with decreasing temperature, an approximate condition for
the appearance of this type of correlation can be obtained from
\Eref{entanglement} in terms of the
longitudinal susceptibility alone. This is given by, $\chi_{z}<N\mathit{l}%
/2kT$. For the specific case that the spin $\mathit{l}=1$, this
condition can
be made precise \cite{japa}. It turns out that whenever, $\chi_{z}%
(T,H=H_{C2})<(9/16)(N/kT)$ entanglement sets in among the spins.

As temperature increases along the line $H=H_{C2}$ the
entanglement measure $E(T)$ also increases. Since the
magnetization for  $H=H_{C2}$ decreases as
$M=N\mathit{l}(1-aT^{3/2})$, the complementarity relation
\Eref{complementarity} implies that,
\begin{equation}
E(T)=1-\frac{kT\chi}{N\mathit{l}}%
\leq1-(1-aT^{3/2})^{2}.
\end{equation}
This can be written as, $E(T)\leq f(T/T_{C})$ where $T_{C}=\left(
1/a\right)  ^{2/3}$. The quantity $a$ is easily calculated and we
get ($\mathit{l}=1$),
\begin{equation}
T_{C}=\frac{1}{\zeta(3/2)^{2/3}}\frac{2\pi\hbar^{2}}{mk_B}\frac{1}{v^{2/3}}.
\end{equation}
This is just the Bose-Einstein condensation temperature of a
system of $N$ bosons. In this expression, $m=(\hbar^{2}/2D)$ is
the mass of the magnons with spin-wave stiffness $D$ and $v=V/N$,
with $N$ the total number of spins in the volume $V$. In our
problem described by \Eref{uzunov} this arises due to the
constraint in the number of modes and that the dominant
contribution for the decay of the magnetization is the Gaussian
one, the interaction $v_0$ being truly irrelevant for this
quantity (see Table \ref{table}). Thus at low temperatures where
the spin-wave approximation holds entanglement scales with the
characteristic temperature $T_{C}$ at the quantum critical point.
This is an interesting feature of this quantum phase transition
associated with  a soft mode. Although the crossover temperature
$T_{\times }\propto|\delta|^{\nu z}$ vanishes at the QCP and
excitations become gapless, the mass of the bosons remain finite
providing an energy scale even at the QCP.

It is useful to explore the analogy of the present magnetic
problem with a true Bose-Einstein condensation of bosonic
particles \cite{bose} to gain insight in the latter problem. The
relevant Bose-Einstein transition in this case is the zero
temperature density-driven transition in a system of interacting
bosons from the incompressible insulating phase to the superfluid
\cite{fisherm,kawa,kawa2}. The control parameter $\delta$ is given
by $\mu-\mu_{C}$ where $\mu_{C}$ is the critical (interaction
dependent) value of the chemical potential. The magnetization in
the magnetic problem corresponds to the number of condensed bosons
and the longitudinal susceptibility to the compressibility defined
as $\kappa=-\partial ^{2}f/\partial\mu^{2}$. The transverse
uniform susceptibility can now be associated with the order
parameter susceptibility of the superfluid \cite{bose} and
diverges for $T\rightarrow0$ at the QCP \cite{comment},
$\mu=\mu_{C}$. The analogy with the magnetic case yields a
criterion for the appearance of entanglement along the quantum
critical trajectory ($\mu=\mu _{C}$, $T\rightarrow0$) that can be
expressed solely in terms of the compressibility. For
($\mathit{l}=1$) this is given by, $\kappa<(9/16)(N/kT)$. The
characteristic temperature which constraints the entanglement
measure is the Bose-Einstein critical temperature of bosons with
density $n(\mu_{C})$. Entanglement in this case implies the
establishment of phase coherence among the particles \cite{bose}.

\section{Conclusions}

We have obtained the thermodynamic properties at the quantum
critical point of magnetic field induced phase transitions using a
scaling approach. We considered first the zero temperature
transition from the saturated paramagnetic phase to a phase with
long range order in the transverse components of the magnetization
with decreasing field. We also presented results for the field
induced transition from the spin liquid to the antiferromagnet.
These transitions are in different universality classes, due to
the distinct values of the dynamic critical exponents. These are
determined by the dispersion relation of the gapless excitations
at the QCP. Since for $3d$ systems and the dynamic exponents
considered here $d_{eff} \ge d_{c}$, the critical exponents
associated with both QCP can be immediately obtained in this case.
Although for $d_{eff} \ge d_c$ the fixed point governing the
quantum phase transition is Gaussian, in both cases, the quartic
term in the action due to spin-wave interaction is dangerously
irrelevant and must be considered. It plays a fundamental role in
determining the critical line and the behavior along the quantum
critical trajectory of the correlation length and susceptibility.

For the problem described by the action \Eref{uzunov}, the zero
temperature exponents remain Gaussian even below $d_c=2$
\cite{uzunov}. However, as discussed below \Eref{tilde} the
behavior along the quantum critical trajectory in the presence of
dangerously irrelevant interactions it is also affected by the
thermal exponents. For this reason we restricted our analysis to
$3d$ systems. Finally, we have investigated entanglement
properties along the critical line as witnessed by the magnetic
susceptibility. We have shown that at the QCP, the Bose-Einstein
condensation temperature provides a well defined energy scale for
the appearance of entanglement among the spins

\ack{ I would like to thank Tatiana Rappoport for useful
discussions. Support from the Brazilian agencies CNPq and FAPERJ
is gratefully acknowledged. Work partially supported by PRONEX/MCT
and FAPERJ/Cientista do Nosso Estado programs.}


\begin{thebibliography}{99}
\bibitem {comment} The nature of the transverse order in the
magnetic problem, either ferro or antiferromagnetic is irrelevant
\cite{kawa} as their QCPs are in the same universality class.
However, it is more useful to carry on the analogy of the
Bose-Einstein problem with the ferromagnet, since in this case,
the uniform transverse susceptibility is the order parameter
susceptibility with well defined scaling behavior at the QCP.
\end{thebibliography}
\end{document}